\documentclass[fleqn,usenatbib]{mnras}
\usepackage{newtxtext,newtxmath}

\usepackage[T1]{fontenc}

\DeclareRobustCommand{\VAN}[3]{#2}
\let\VANthebibliography\thebibliography
\def\thebibliography{\DeclareRobustCommand{\VAN}[3]{##3}\VANthebibliography}

\usepackage{graphicx}	
\usepackage{amsmath}	
\usepackage{amssymb}	

\title[A Fresh Look at AGN SED Fitting]{A Fresh Look at AGN Spectral Energy Distribution Fitting with the XMM-SERVS AGN Sample}

\author[A. Marshall]{
Adam Marshall$^{1}$\thanks{E-mail: alm99@cam.ac.uk}, Matthew W. Auger-Williams$^{1,3}$, Manda Banerji$^{2}$\, Roberto Maiolino$^{3,4,5}$, Rebecca Bowler$^{6,7}$\
\\
$^{1}$Institute of Astronomy, University of Cambridge, Madingley Road, Cambridge CB3 0HA, UK\\
$^{2}$School of Physics and Astronomy, University of Southampton, Highfield Campus, Southampton SO17 1BJ, UK\\
$^{3}$Kavli Institute for Cosmology, University of Cambridge, Cambridge CB3 0HE, UK\\
$^{4}$Cavendish Laboratory, University of Cambridge, Cambridge CB3 0HE, UK\\
$^{5}$Department of Physics and Astronomy, University College London, Gower Street, London WC1E 6BT, UK\\
$^{6}$Astrophysics, The Denys Wilkinson Building, University of Oxford, Keble Road, Oxford, OX1 3RH, UK\\
$^{7}$Jodrell Bank Centre for Astrophysics, Department of Physics and Astronomy, School of Natural Sciences, The University of Manchester, Manchester, M13 9PL, UK\\
}

\date{Accepted XXX. Received YYY; in original form ZZZ}

\pubyear{2022}

\begin{document}
\label{firstpage}
\pagerange{\pageref{firstpage}--\pageref{lastpage}}
\maketitle
\begin{abstract}
We perform spectral energy distribution (SED) fitting to 711 luminous X-ray AGN at $0.7 < z < 4.5$ using 10-bands of optical and infra-red photometric data for objects within XMM-SERVS.  This fitting provided 510 reliable (reduced $\chi ^2 < 3$) inferences on AGN and host galaxy properties. The AGN optical (3000\AA) luminosity inferred from SED-fitting is found to correlate with the measured X-ray (2-10 keV) luminosity, in good agreement with previous work. Using X-ray hardness as a proxy for AGN obscuration, we also study the differences in the host galaxy properties of obscured and unobscured AGN. Both populations have consistent stellar masses (log$_{10}(M_*/M_{\odot})$ = 10.88 $\pm0.09M_\odot$ and log$_{10}(M_*/M_{\odot})$ = 10.8 $\pm0.1M_\odot$ for unobscured and obscured AGN respectively). We also find evidence for varying AGN emission line properties from a standard AGN template in 18.8\% of the sample with a reduced $\chi^2 < 3$ where the inclusion of an additional emission line strength free parameter was found to improve the quality of the fit. Comparison of these fits to SDSS spectra showed that emission line properties inferred from broadband photometry were consistent with the results from spectroscopy for 91\% of objects. We find that the presence of weaker, more blueshifted emission lines as inferred from the SED fits are associated with more negative values of $\alpha_{ox}$. While the correlation between the hardness of the ionising SED and the emission line properties has been known for some time, we are able to derive this correlation purely from broadband photometry.

\end{abstract}

\begin{keywords}
quasars: general --  quasars: emission lines -- galaxies: active
\end{keywords}


\section{Introduction} 
Observations of massive galaxies suggest that most contain supermassive black-holes (SMBHs) at their centres with masses greater than $10^6M_\odot$ \citep{magorrian}. Despite the difference in relative scale between these black holes and their host galaxies, our current paradigm of galaxy formation suggests that galaxies and their supermassive black holes co-evolve \citep{croton, Sijacki_2015} with feedback processes \citep{fabian_2012} regulating the growth of both the supermassive black hole and the host. Studying the link between galaxies and black holes requires us to measure physical properties associated with both, and has been attempted in numerous previous works (e.g. \citealt{merloni, bongiorno_2012,azadi,Circosta,mountrichas,Pouliasis}).

Both host galaxy and black hole properties can potentially be constrained through analysis of spectra for galaxies with Active Galactic Nuclei (AGN). However, since the advent of large multi-wavelength galaxy surveys, SED-fitting has been routinely employed in order to infer photometric redshifts (e.g. \citealt{salvato_2019}) as well as galaxy and AGN physical properties for statistical samples (e.g. see \citet{walcher_2011, johnson_2021, thorne_2022} for a review of commonly used methods). At rest-frame ultraviolet (UV) and optical wavelengths much of the focus in SED-fitting has been on improving stellar population synthesis models to more accurately represent the emission from stars in galaxies (e.g. \citet{kriek_2013, conroy_2013} and references therein). Bayesian techniques are also increasingly being employed to constrain SED properties (e.g. \citealt{rivera}) with greater awareness of some of the potential pitfalls of interpreting simple maximum likelihood estimates \citep{mountrichas}.

Within the galaxy SED fitting community, there is a recognition that SED-fit parameters can have complex degeneracies in the multi-dimensional fitting space (e.g. \citealt{lower_2020}), and underestimating the real inherent uncertainties in these fits has potential consequences for what we can conclude from them about galaxy formation and evolution (e.g. \citealt{Curtis-Lake_2021}). For high-redshift galaxies, the realisation that emission lines can contribute significant flux in some passbands and therefore influence the best-fit SED model has revolutionised our understanding of the results from SED-fitting (e.g. \cite{Schaerer_2009, debarros_2013, smit_2014}. Contemporaneously to these advances, there have also been notable developments in SED-fitting techniques that use a self-consistent approach to simultaneously model the ultraviolet through far infrared emission from galaxies -- e.g. CIGALE \citep{cigale}, MAGPHYS \citep{daCunha_2012}. Large survey datasets that extend into the infrared, such as surveys conducted with the \textit{Herschel} Space Telescope \citep{pilbratt_2010}, have driven these improvements to be able to model the cool dust emission from galaxies. AGN components are more commonly incorporated into SED modelling codes that cover an extensive wavelength range primarily because the multi-wavelength data can help break some of the degeneracies between the AGN and host galaxy parameters \citep{rivera}.

There are two broad types of AGN templates commonly employed in SED-fitting codes: (i) empirically derived templates (e.g. \citealt{polletta_2007, richards_2006}) based on observations of known AGN and (ii) theoretical templates produced using radiative transfer models (e.g. \citealt{fritz_2006, stalevski_2016}). The empirical templates, while providing a relatively simple parametrisation of the AGN emission, may not be representative of all AGN. The theoretical SEDs offer more flexibility to model diverse AGN emission but at the expense of a very large number of free parameters, many of which are difficult to constrain using broadband photometric data alone. Moreover, none of these templates have, as yet, assessed critically the effect of emission and absorption features to the broadband SED fitting, in a way analogous to what has been done for high-redshift galaxies.

The rapid advances in precision imaging datasets in the optical and near infrared -- e.g. the Dark Energy Survey (DES;\citealt{abbott_2021}), HyperSuprimeCam \citep{HSC}, the upcoming Vera C. Rubin Observatory Legacy Survey of Space Time (LSST) \citep{lsst} and \textit{Euclid} \citep{euclid} - means optical and infrared surveys are already far surpassing the flux limits achievable over a wider wavelength range. In the context of jointly studying AGN and host galaxy emission, this necessitates the development of parallel SED-fitting approaches that attempt to model both the AGN and host galaxy over a more limited wavelength range, with a relatively small number of free parameters and to the resolution required to match current and future large sky surveys. New wide-field spectroscopic surveys such as 4MOST \citep{4most} and VLT-MOONS \citep{maiolino_2020} will also use SED-fitting to broadband photometry as the basis for their AGN target selection. Thus, in light of these new surveys, it is timely to critically assess how well current SED-fitting methods are able to jointly constrain galaxy and AGN properties. That is the aim of this study.

In the present work we make use of a new empirical quasar SED from \cite{temple} (T21 hereafter) to model AGN emission. The model differs from previous AGN SED models in its aim to accurately reproduce the average colours of unobscured AGN over a more limited wavelength range (rest frame 912\AA - 3$\mu$m) using only a small number of free parameters. Notable improvements in the T21 model relative to previous work include a more accurate determination of the contamination from the host galaxy to the quasar continuum, as well as a thorough treatment of the effect of broad emission lines on the broadband quasar colours. The former improvement allows a more robust determination of the `pure AGN' emission, therefore consequently providing more reliable host galaxy properties. Even in the case of bright AGN at $z > 2$, host galaxy contribution was found to account for $>5\%$ of the flux of the total SED \citep{temple}. As such, not accounting for such a contribution could bias host galaxy stellar mass estimates. Additionally, the \citet{temple} model also encompasses the full range of possible emission line properties seen in quasar spectra, from weak, highly blueshifted lines through to high equivalent width, symmetrical lines. Their analysis found that emission lines could affect AGN photometric colours by 0.1 magnitudes or more in some cases, a difference that could easily be measured within modern photometric surveys with typical uncertainties of $\sim$0.05 mag.
 
In this paper we combine the T21 AGN SED model with galaxy templates from \cite{fsps1,fsps2} to fit the observed optical to infrared SEDs of a sample of X-ray selected AGN. We make use of a Bayesian SED-fitting technique and Markov Chain Monte Carlo (MCMC) sampling to fully explore the AGN+host galaxy parameter space. Our focus is on relatively luminous and distant ($z>0.7$) AGN. We find, consistent with many previous studies, that host galaxy parameters such as age, dust extinction and star formation rate are poorly constrained using  only optical and near infrared photometry \citep{ciesla}. We therefore focus primarily on the AGN properties such as luminosity, obscuration and emission line properties, and their link to the host galaxy stellar mass, which can be measured more reliably, provided that photometry sampling rest-frame wavelengths of $\sim1\mu$m is available. This wavelength is where the AGN emission reaches a minimum, thereby enhancing the contrast between an old galaxy stellar population and the AGN \citep{merloni, bongiorno_2012}. A companion paper to this one (Marshall et al. in prep) presents a more detailed comparison of how the choice of AGN template used in SED fitting can affect host galaxy stellar mass estimates.

This paper is structured as follows: first we describe the dataset in Section \ref{data}. The SED fitting method is introduced in Section \ref{method}, and the main results from our study are summarised in Section \ref{results} with our inferences on our AGN luminosities in the optical, infrared and X-ray. In Section \ref{mstar} we discuss how these luminosities relate to the stellar mass of the AGN host galaxy. We then look in Section \ref{emtype} at the inferred emission line properties for our objects, in comparison to the actual emission lines seen in SDSS spectra. Finally we  look at how our inferred emission lines relate to the X-ray to UV slope ($\alpha_{ox}$). Throughout the paper, all magnitudes used are on the AB system. We adopt a  $\Lambda$CDM cosmology with $\Omega_m$=0.3, $\Omega_\Lambda$=0.7, $H_0$ = $70 kms^{-1}Mpc^{-1}$. 

\begin{table*}
 \caption{Model parameters for our galaxy and AGN SED templates.}
 \label{tab:para}
 \begin{tabular}{lll}
  \hline
  Parameter & Range & Additional notes \\
  \hline
  \textbf{Galaxy parameters}\\ 
  Star formation rate e-folding time ($\tau$) & 0.08 $\leq$ $\tau$\ $\leq$ 25 Gyr  \\
  Effective v-band Optical depth $\tau$V & $10^{-4}$ $\leq$ $\tau$V $\leq$ 2 & All galaxy parameters\\
  Age (Years)  & $3.2 x 10^{5}$ $\leq$ Age $\leq$ $1.4 x 10^{10} $ & have priors flat in log space\\
  Stellar Mass (log$_{10}(M_*/M_{\odot}$) & $10^{7}$ $\leq M_*$ $\leq$ $10^{13} M\odot$\\
  \\ 
  \textbf{AGN parameters}\\
  Quasar Luminosity at 3000{\AA} & $10^{40}$ $\leq$ L $\leq$ $10^{50}$ ergs$^{-1}$ & Priors flat in log space\\
  Quasar Reddening E(B-V)& -0.2 $\leq$ E(B-V) $\leq$ 2\\
  Hot dust Luminosity at 3$\mu$m & $10^{30}$ $\leq$ L $\leq$ $10^{49}$ ergs$^{-1}$ & Priors flat in log space\\
  Emission line strength & -2 $\leq$emline type $\leq$ 3\\
  \hline
 \end{tabular}  
\end{table*}

\section{Data} \label{data}
The parent sample of X-ray selected AGN considered in this work originates from the XMM-SERVS survey \citep{XMM}, which provides 5242 AGN candidates over 5.3 deg$^2$ of the XMM-Large Scale Structure (XMM-LSS) survey region with an X-ray survey integration time of $\approx$ 50ks. The advantage of X-ray selection is the ability of the hard X-ray photons to penetrate significant dust and gas columns, thereby making identification of heavily obscured AGN possible. X-ray hardness ratios can also be used to separate AGN based on their levels of obscuration, as will be discussed further in Section \ref{agnlum}. 

The XMM-Newton survey region overlaps with several other optical and infrared multi-wavelength datasets including the \textit{grizY} bands of HyperSuprimeCam Deep \citep{HSC}, \textit{ZYJHKs} bands from VISTA VIDEO \citep{VIDEO}, and the 3.6$\mu$m and 4.5$\mu$m bands of the Spitzer Extragalactic Wide-area IR Survey (SWIRE; \citealt{SWIRE}). \citet{XMM} have identified optical and infrared counterparts to the X-ray sources using a likelihood ratio method, with 93\% of the X-ray sources found to have reliable counterparts in HSC and VIDEO. In addition, 82\% of the X-ray sources have reliable mid infrared counterparts in either the 3.6$\mu$m or 4.5$\mu$m imaging from SWIRE. Whilst VISTA VIDEO also provides Z- and Y-band photometry, we limit our analysis to only using the higher precision, and more complete within the survey region, HSC z-and Y-bands. Our multi-wavelength data therefore constitutes 10 bands of photometry ranging from the optical $g$-band through to 4.5$\mu$m.

We update the optical and near infrared photometry presented in \citet{XMM} by considering the latest data releases from both the HSC PDR2 \citep{HSCDR2} and VISTA VIDEO surveys \citep{bowler}. We use aperture photometry with a 3$\arcsec$ diameter aperture and an aperture correction. To avoid over-fitting due to unrealistically small uncertainties, a minimum uncertainty limit of 5$\%$ of the 3$\arcsec$ aperture flux was placed on each band. We further restrict our sample to only those sources with reliable spectroscopic redshifts -- 1314 in total. We choose to limit to objects with spectroscopic redshift, as the  inclusion of redshift as an additional free parameter introduces further degeneracies. Visual inspection of the HSC \textit{gri} images suggests that a large number of the lowest redshift sources are galaxies hosting low-luminosity AGN. As the primary goal of this work is to investigate the spectral energy distributions of high-redshift, high luminosity AGN, we therefore place a redshift cut of $z>0.7$ on our sample. This gives a total of 774 spectroscopically confirmed X-ray selected AGN. Finally, to avoid object blends affecting the optical and infrared photometry, we further remove any AGN with a neighbour in the full HSC DR2 catalogue that is $<$2$\arcsec$ from the AGN itself. This leads to a final sample of 711 spectroscopically confirmed, X-ray selected AGN, whose spectral energy distributions are studied in detail in this work.

\section{Method} \label{method}
In order to model 10-bands of photometry within the optical and infrared, we incorporate three luminous components which are combined to create a total model SED. These components are the AGN accretion disk and broad line emission, the hot dust emission from the AGN, and the AGN host galaxy (including stellar light and nebular emission). Due to the chosen redshift and wavelength range of the data, we do not include the contribution of cooler dust, which may provide significant flux at longer wavelengths. We describe the galaxy and AGN templates below.

\subsection{Galaxy Template}
Galaxy templates were produced using the Flexible Stellar Population Synthesis (FSPS) code \citep{fsps1,fsps2}, including nebular emission lines. Composite stellar populations (CSPs) were produced assuming an initial mass function defined by \cite{chabrier} and solar metallicity. We assume exponentially declining star-formation histories with a range of e-folding times and range of ages as detailed in Table \ref{tab:para}. Dust extinction is applied to these templates, assuming the \cite{calzetti} attenuation curve, for a range of optical depths. Finally, the normalisation of the galaxy template provides an additional free parameter corresponding to the stellar mass of the galaxy. This provided  a total of four free parameters associated with the AGN host galaxy.

\begin{figure*}+
    \centering
    \includegraphics[width=\linewidth]{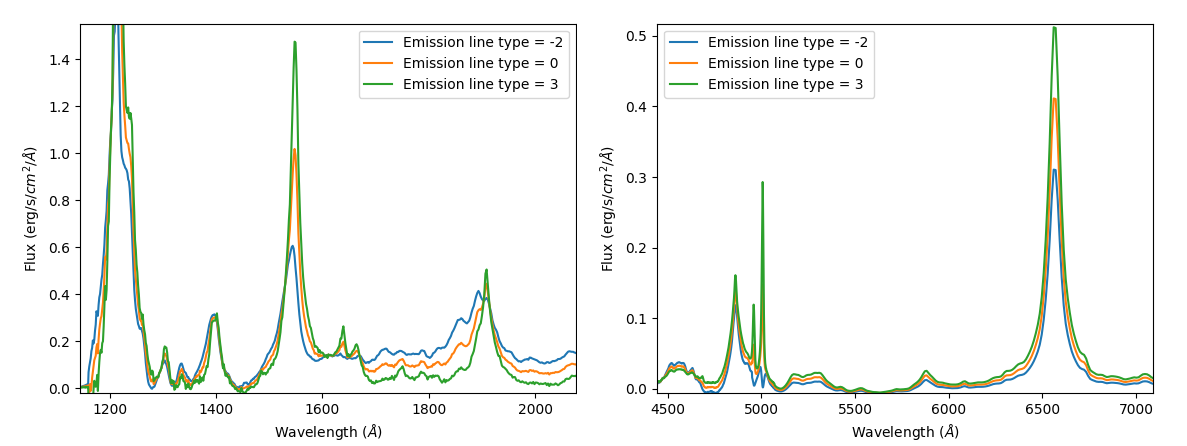}
    \caption{The range of emission lines properties available in the T21 AGN template spectrum, as parametrised by the emline\char`_type. Negative values correspond to weaker, more highly blueshifted emission lines, whereas positive values correspond to stronger, more symmetric emission lines. An emline\char`_type of 0 corresponds to the average emission line properties for an SDSS quasar at z = 2 with an absolute magnitude $M_i$ = -27.}
    \label{fig:emline_comp}
\end{figure*}

\subsection{AGN Template}
\label{AGN_properties}
The AGN SED model is described in detail in \citet{temple} and includes contributions from the quasar accretion disk, broad and narrow emission lines, and the hottest component of the dusty torus emitting at close to the sublimation temperature. The model is empirically derived using the spectra of known quasars in the Sloan Digital Sky Survey. Key improvements of this model relative to previous quasar SED models include a better understanding of both the evolution of quasar emission line properties and the host galaxy contribution as a function of luminosity and redshift (see \citealt{temple} for more details). The free parameters in the model include the quasar reddening E(B-V), assuming an empirically derived extinction curve \citep{temple}. In the case of some of our fits, the emission line properties are also allowed to vary between a range of values (-2 and 3), as shown in Fig. \ref{fig:emline_comp}. We are able to gain information on the relative blueshift of AGN emission lines from photometry due to the intrinsic link of the emission line morphology to emission line strength \citep{temple}. These emission line property values are analogous to the emission line range shown in the top of Fig. 3 in \cite{Richards_2021}. Negative values of the emission line type correspond to highly blueshifted, weak emission lines similar to objects in the bottom right of \cite{Richards_2021} Fig. 3, whereas positive values of the emission line property correspond to stronger, more symmetrical emission lines, similar to objects in the top left of that figure. The normalisation of the rest-frame UV to optical SED also provides a free parameter that provides a measure of the AGN luminosity. Finally, the hot dust component is modelled as a blackbody with a fixed temperature of 1236K, whose normalisation provides us with the hot dust luminosity at 3$\mu$m. Cooler components of the dust do not affect the observed-frame colours at $\lambda < $4.5$\mu$m given our redshift cut of $z>0.7$, and thus are not considered in our models. 

\subsection{MCMC Fitting Algorithm} \label{init}
We conduct a Bayesian MCMC SED fit over the parameter space summarised in Table \ref{tab:para}. Given the diversity of AGN in our sample, we perform four different SED fits, furthermore referred to as model families, and then select the best model for each object, as detailed in Section \ref{bestfd}. The four different MCMC runs correspond to (i) a AGN+GAL SED with the AGN emission properties fixed to the average seen in AGN at z = 2 and an average absolute magnitude $M_i$ = -27 \citep{temple}; (ii) a AGN+GAL SED with variable emission lines properties in the AGN component, between the limits shown in Fig. \ref{fig:emline_comp}; (iii) a galaxy-only SED; and (iv) a AGN-only SED with the same variable emission line properties as used in model family (ii). The latter two runs allow for fitting of objects where the optical to infrared emission is entirely dominated by either the galaxy or the AGN component. For each run, the MCMC algorithm utilised allows for the calculation of a posterior probability, P($\rho$|data) for a specified set of model parameters, $\rho$, via the Bayesian relation:

\begin{equation}
    P(\rho|data) = \dfrac{P(data|\rho)P(\rho)}{P(data)}
\end{equation}

\noindent where P(data) is a normalisation term, and P($\rho$) is a function containing information associated with any prior knowledge of the expected value of each free parameter. For each SED model, the prior information, P($\rho$), for all parameters of the model is a flat distribution. In the cases of $\tau$, $\tau_v$, age and mass for the galaxy templates, the AGN luminosity at 3000\AA\ , and the hot dust luminosity at $3\mu$m, the prior distribution is flat in logarithmic space (Table \ref{tab:para}). P(data|$\rho$) is the likelihood of observing the data given the chosen model parameters $\rho$. For the inferences associated with a specific band, $m_{model_i}$($\rho$), produced from the parameter values, ($\rho$), along with the observed magnitudes ($m_{obs}$) and their associated uncertainties, ($\sigma$), this is calculated within the fitting code using the relation:

\begin{equation}
    P(m_{obs}|\rho)  \propto  \prod_i^n exp\bigg(-\dfrac{[m_{obs_i} - m_{model_i}(\rho)]^2}{2\sigma^2_i}\bigg)
\end{equation}

We anticipate that the posterior will often be multi-modal, with some of the modes having negligible likelihood. We therefore utilise parallel-tempering ensemble MCMC from the package \textsc{emcee}\footnote{Made using emcee Version 2.2.1} \citep{emcee}, with additional temperatures employed to reduce the chance of walkers remaining stuck in local, low-level maxima. Walker positions were initially randomised within the prior space. A second single-temperature run was performed with walker positions starting at the final positions of the lowest temperature walkers from the previous run. We use the samples from the the second run in further analysis, as we expect samples of the secondary run to be close to the global likelihood maximum and be free of low level local maxima.

\begin{figure*}
    \centering
    \includegraphics[width=\textwidth]{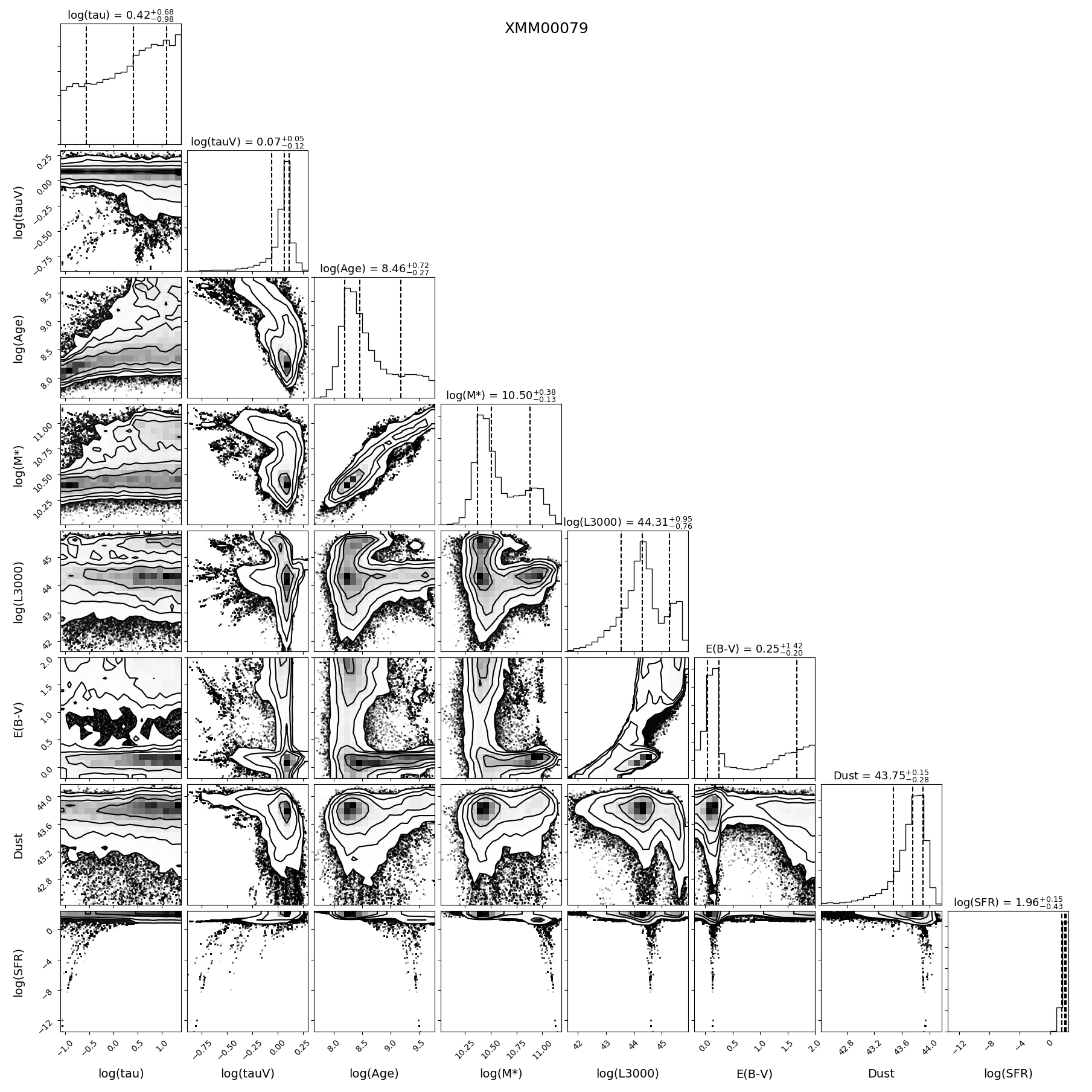}
    \caption{An example of the marginalised one- and two dimensional posterior distributions for an unobscured (HR = -0.67) AGN at z = 0.93. In this example, we can see the degeneracies between a young, low stellar mass, star forming galaxy and an older, redder galaxy with a higher luminosity AGN. This can be seen in the bimodalities of the host galaxy stellar mass and age, and AGN optical luminosity and extinction.}
    \label{fig:corner}
\end{figure*} 

\subsection{Selecting the best model family} \label{bestfd}
The methods described in Section \ref{init} provide four model families for each of our 711 objects. These include the two AGN+GAL fits, one with emission line properties as a free parameter and the other fixed to the average emission line properties for SDSS AGN derived by \citet{temple}, as well as the AGN and galaxy-only fits. All four model families were run for each object. For each run we calculate the Akaike Information Criterion (AIC) using the maximum likelihood, $L$, and the number of free parameters, $k$, as:

\begin{equation}
    AIC = 2k - 2log(L)
\end{equation}

\noindent where the model family with the lowest AIC is considered our best-fit. The benefit of using the AIC to determine which model family best represents the data for each object is that increasingly complex models with a larger number of additional free parameters are penalised in the AIC. We can therefore, for example, determine if allowing the emission line properties to vary is truly improving the quality of the fit significantly enough to justify the increased model complexity. In order to confirm this as a valid method of model family selection, we looked at the objects where the AIC of different model families gave similar values. Of these, only 13 objects were found to have a galaxy-only and QSO+GAL AIC difference < 1. Whilst it may be the case that, for some of these objects, further information on AGN properties could have been inferred, the sample of objects is small, and their inclusion does not have a significant effect on the total distribution of samples shown within the results of this work.

Of the 711 objects, 438 (61.6\%) were best fit with the AGN+GAL model fixed to the average emission line properties in T21, 197 (27.7\%) preferred the AGN+GAL model family with varying emission line properties (see also Section \ref{emtype}), 72 (10.1\%) preferred the GAL only model family and are the most obscured/low-luminosity AGN where the optical to infrared emission is completely dominated by the host galaxy. Finally, only 4 (0.6\%) preferred the AGN only model family. Within some sections of our analysis, we compare our inferred AGN and host galaxy properties. Therefore, to maintain consistency in the sample used throughout our analysis, we do not include inferences from the objects that preferred galaxy-only and AGN-only model families.

\begin{figure*}
    \centering
    \includegraphics[width=\textwidth]{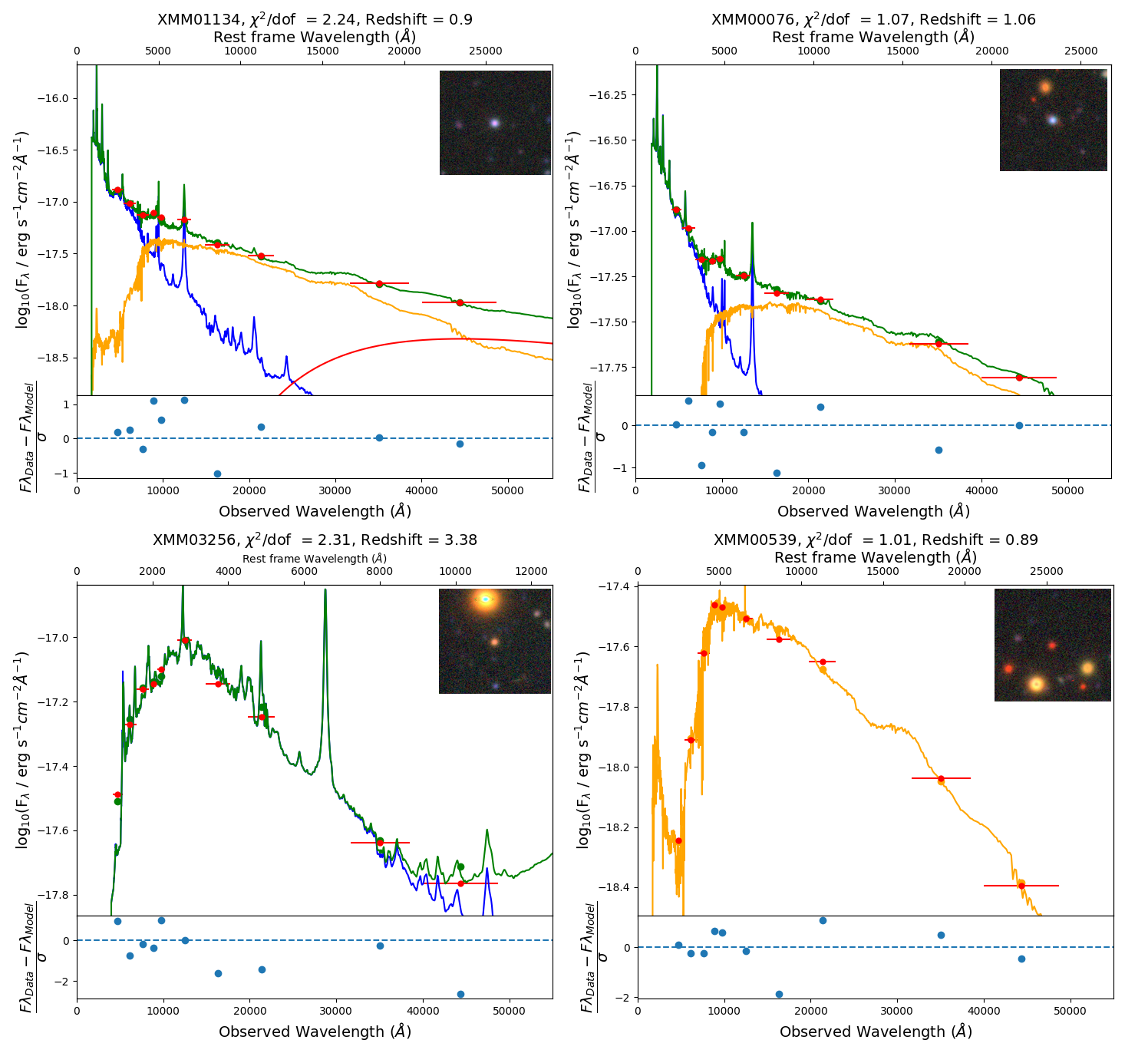}
    \caption{Example fits for the four model families. The top figures show examples of fits with both the galaxy and AGN components, one with a free emission line (left), and the other with emission line properties fixed to the average seen in an SDSS quasar at z=2 AGN, with an average absolute magnitude $M_i$ = -27. The bottom figures show fits where either a AGN-only (left) or Galaxy-only (right) fit is preferred. In each case, the total maximum likelihood spectrum is shown in green. The blue, orange and red spectra represent the AGN, galaxy and AGN hot dust components respectively. The observed photometric data is shown as red points. In each corner is the HSC DR-2 \emph{gri} colour image of the object.}
    \label{fig:comb_sed}
\end{figure*}
After the MCMC fitting had been run for all 711 objects within the sample, and each object had been assigned a preferred model family, a visual inspection of the marginalised one- and two-dimensional posteriors for each run was performed. An example of these posteriors is shown in Fig. \ref{fig:corner} where the age, stellar mass, and E(B-V) are bimodal. We find that the age is in general poorly constrained for the majority of our sources, as is often the case when fitting spectral energy distribution models to broadband photometry without the use of UV bands \citep{ciesla}. Whilst the example shown in Fig. \ref{fig:corner} is more bimodal than the majority of our objects, the degeneracies visible highlight a typical issue for composite host galaxy and AGN SED fitting: a dust-reddened quasar residing in a young star-forming galaxy is often degenerate with an unobscured AGN in an older galaxy. The posterior distributions shown in E(B-V) of Fig. \ref{fig:corner} are also often non-Gaussian with visible tails. For completeness, when comparing AGN and host galaxy properties, we include within our figures both the distribution of all of the MCMC inferences from each object, along with the corresponding median solutions inferred for each property. Both our median and all MCMC inferences show similar distributions for the AGN and host galaxy properties that we investigate in this paper.

Before summarising the main results from our work, we make one final cut to the sample to remove objects where the highest likelihood solution still provides a reduced $\chi^2$ $>$ 3. This cut was chosen to be fairly liberal to account for the possible under-prediction of uncertainties in the photometric data, whilst still removing objects with poor fits based on a visual inspection. This leaves a final sample of 510 objects, which is comparable in size to previous work by e.g. \cite{Lanzuisi} in the COSMOS field. Of these 510 objects, 437 preferred a AGN+GAL model family, either with emission line properties that are allowed to vary within the fitting run (18.8\%), or fixed to the average seen in AGN at z = 2 (81.2\%) and an average absolute magnitude $M_i$ = -27. Of the remaining objects that had a reduced $\chi ^2$ < 3, 71 instead preferred a galaxy-only model family, and 2 an AGN-only model family. 

In order to study the difference between obscured and unobscured AGN within our sample, we also separated our 437 AGN+GAL model family objects using their measured X-ray hardness ratio provided by \cite{XMM}. A hardness ratio of -0.2 was selected as the threshold, with hardness ratio values below and above this value corresponding to unobscured and obscured objects respectively \citep{hasinger}. This resulted in 316 unobscured AGN ($72\%$ of the AGN+GAL sample), and 121 obscured AGN (28\% of the AGN+GAL sample). Of the further 71 objects that were found to have best-fit templates that were dominated by the galaxy component, all but 18 only provided upper limits on either the hard- or soft X-ray detections. Of these, 13/18 (72$\%$) have hardness ratios $>$ -0.2. We might expect our objects dominated by the galaxy in the optical and infrared region to also be more heavily obscured in the X-ray, which appears to be the case, especially when we consider that the majority of our galaxy sample appear to be so obscured that a definitive measurement was not possible within one of the X-ray bands.

\begin{figure*}
 \centering
    \includegraphics[width=\textwidth]{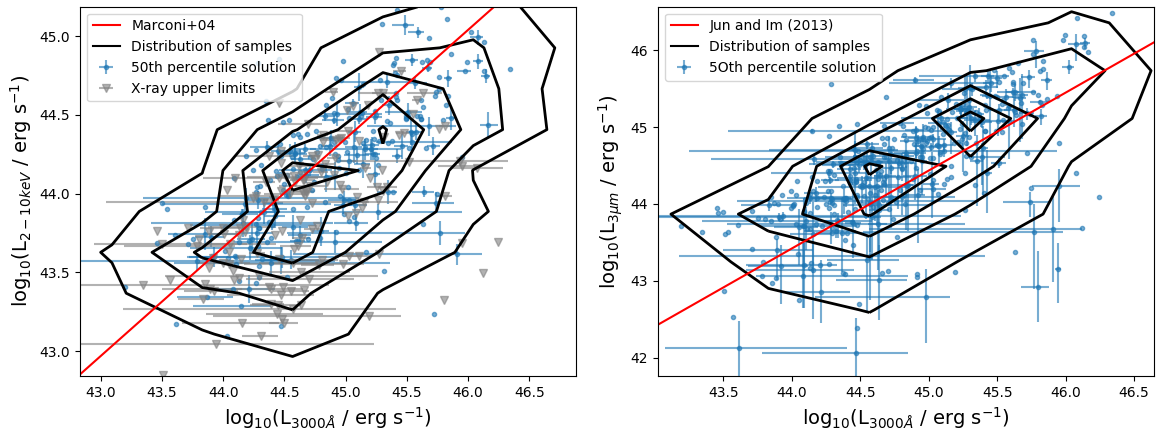}
    \caption{XMM-Newton X-ray luminosity vs AGN optical luminosity at 3000\AA\, with the relation derived from \protect\cite{marconi} (left) and the 3000\AA\ optical luminosity vs hot dust luminosity at 3$\mu$m with the relation from \protect\cite{jun_im} (right). The blue data points show the median solutions and uncertainties for these objects, and the contours show the combined inferences from our MCMC analysis. For clarity, uncertainties are limited to a third of the points shown. In the case where X-ray luminosity is plotted, objects where log$_{10}(L_{2-10keV})$ is an upper limit are shown in grey.}
    \label{fig:types_comb}
\end{figure*}

\section{Results} \label{results}
In Fig. \ref{fig:comb_sed} we show example SED fits from the four model families of objects: AGN dominated (bottom-left), galaxy dominated (bottom-right), AGN+GAL with the AGN emission lines fixed to the average emission line template from T21 (top-right), and AGN+GAL with variable emission line properties (top-left). Each example also meets our reduced $\chi^2$ < 3 cut. We find that the majority of SEDs that are best fit by a AGN+GAL model family have a relatively unobscured AGN dominating the near-UV with old stellar populations from the host galaxy providing flux at longer wavelengths, as shown in the upper two panels of Fig. \ref{fig:comb_sed}. The hot dust emission from the AGN only contributes beyond a rest-frame wavelength of $\sim$1$\mu$m. Our inferences from the SED fits are broadly consistent with the HSC $gri$ colour composite images shown in the inset panels. For example, in the top-right source XMM00076, the HSC image shows a blue point source surrounded by extended red emission. This is consistent with the bluer flux being dominated by the quasar and the redder flux having more significant contributions from the extended host galaxy. The upper left hand panel shows the best fit for XMM01134 using non-standard emission lines, preferring stronger, more symmetric lines compared to the average SDSS quasar. These lines appear to contribute significantly to the broadband flux in the Y and J bands. The subset of AGN fit with atypical emission lines will be discussed further in Section \ref{emtype}.

The bottom-right panel of Fig. \ref{fig:comb_sed} shows an example of one of the 71 objects that preferred a galaxy-only fit based on our AIC selection. These subset of objects are still actively accreting based on the detection of significant X-ray emission, but the AGN contribution is obscured across the entire optical and infrared region of the spectrum. The average redshift of the galaxy-dominated objects is $z = 0.89$, significantly lower than the AGN+GAL class of objects, which have an average redshift of $z = 1.53$. We consider the difference between the HSC $g$-band cModel and PSF magnitudes provided by HSC DR2 \citep{HSCDR2} as a simple measure of extendedness at the bluer wavelengths and find that the AGN+GAL class of objects have an average gcModel-gPSF of -0.12 compared to -0.63 for the galaxy dominated class of sources. This supports that the galaxy dominated sources are indeed more extended in the HSC images. Our galaxy-dominated sample therefore accounts for a subset of AGN with low X-ray luminosities that are only observed within this sample due to being relatively nearby. This results in domination from the host galaxy across the observed wavelength range.

\begin{figure*}
    \centering
    \includegraphics[width=\textwidth]{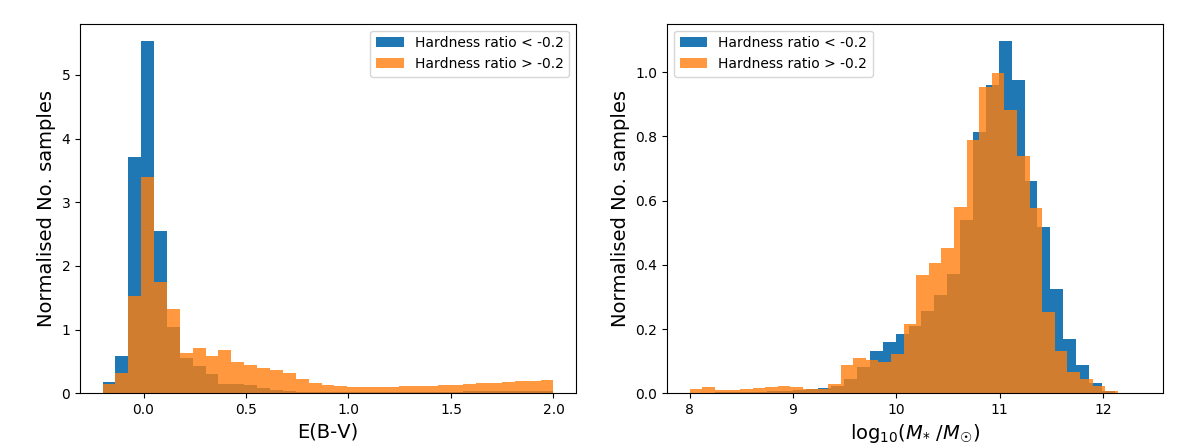}
    \caption{Histograms showing the distribution of E(B-V) (left) and host galaxy stellar mass (right) MCMC inferences, separated by the measured X-ray hardness ratio. This ratio acts as a proxy for the AGN type. We have assumed values above and below HR = -0.2 to represent obscured and unobscured AGN respectively. The obscured AGN show a tail to higher extinction values when compared to the unobscured sample.}
    \label{fig:ebvhist}
\end{figure*}

\subsection{AGN Luminosity \& Obscuration} \label{agnlum}

We now use our SED fits to further explore the inferred AGN luminosities and obscuration for the XMM-SERVS spectroscopic sample. As stated in Section \ref{bestfd}, due to the non-Gaussian nature of many of the 1-D marginalised probability distributions, we choose to show both the median solutions with uncertainties for each object, along with the contours corresponding to every MCMC inference in the posterior distribution for the 437 well-fit AGN+GAL model family objects. The left side of Fig. \ref{fig:types_comb} shows the 2-10keV X-ray luminosity, log$_{10}(L_{2-10keV}$), calculated from the X-ray flux in the catalogue from \citet{XMM}, as a function of the extinction corrected 3000\AA\ luminosity, log$_{10}(L_{3000\AA}$), inferred from the AGN contribution to the SED. The contours represent the density of MCMC inferences that probe the full posterior distributions of the 3000\AA\, luminosity. The straight line is the relation derived from Eq. 21 in \citet{marconi}:

\begin{equation}
log_{10}(L_{2-10keV}) =  0.69 log_{10}(L_{3000\AA}) + 13.3
\end{equation}

\noindent We see that the distribution of 3000\AA\ luminosities is consistent with the relation, with a 1$\sigma$ scatter of $\pm0.4$ dex around the \citet{marconi} line for all of the sources in the sample. As the X-ray luminosities are independent of our SED fits, agreement of the AGN optical luminosity with the predictions from the \citet{marconi} relation serves as a useful validation of our method.

To further understand the relationships between AGN luminosities, we also look at our inferences of the luminosity of the hot dust surrounding the SMBH. The right panel of Fig. \ref{fig:types_comb} shows the distribution of optical luminosity at 3000\AA\ with the AGN hot dust luminosity at 3$\mu m$, log$_{10}(L_{3\mu m}$), for the same sample of X-ray AGN. We also show the empirically derived relation from \cite{jun_im}:

\begin{equation}
    log_{10}(L_{2.3\mu m}) = (1.014 \pm 0.002)\ log_{10}(L_{0.51\mu m}) - (0.655 \pm 0.076)
\end{equation}

The \citet{jun_im} relation compares the hot dust luminosity at 2.3$\mu m$ and the AGN luminosity at 5100\AA. We therefore convert these values to the optical luminosity at 3000\AA\, and the hot dust luminosity at 3$\mu$m. As we are not changing the shape of the unreddened T21 AGN template and dust blackbody, we can convert the luminosities as:

\begin{equation}
    log_{10}(L_{2.3\mu m}) = log_{10}(L_{3.0\mu m}) - 0.058
\end{equation}

\begin{equation}
    log_{10}(L_{5100\AA}) = log_{10}(L_{3000\AA}) - 0.592
\end{equation}

Even after making this conversion, our points are offset by -0.23 dex from the \cite{jun_im} relation, with our SED fits inferring higher 3$\mu m$ luminosities for a given 3000\AA\, luminosity. A key difference between our analysis and that of \citet{jun_im} is the incorporation of the new quasar template from \citet{temple}, which is believed to provide a more accurate representation of the intrinsic quasar SED without contamination from the host galaxy emission. The work by \cite{jun_im} modelled the AGN SED as a power law continuum and hot dust blackbody emission, but did not include the contribution of emission lines. It is therefore possible that the observed discrepancy is due to the intrinsic difference in the AGN SED templates used. The effects of the inclusion of emission lines within AGN templates during SED fitting requires further study, and will form the basis of an upcoming companion paper to this one (Marshall et al. in prep).

Separating out AGN by hardness ratio we found that the distributions of unobscured (HR < -0.2) and obscured (HR $>$ -0.2) objects appear similar, with the main difference being a larger spread on the distribution of inferences for the obscured sample of AGN. The 1$\sigma$\ scatter compared to the optical and X-ray luminosity relation from \cite{marconi} increases from $\pm0.4$ dex to $\pm0.5$ dex from the X-ray unobscured to the obscured AGN. Whilst the AGN luminosity distributions are relatively similar, differences in the measured optical extinction between X-ray unobscured and X-ray obscured AGN are more significant, as can be seen on the left of Fig. \ref{fig:ebvhist}. The histogram of obscured E(B-V) inferences has a tail extending to higher E(B-V) values, whereas the inferences from the sample of unobscured objects are tightly peaked at E(B-V) = 0.02. This difference is expected, as, assuming the unified theory of AGN, flux from obscured AGN passes through a larger amount of dust, leading to greater extinction. If we assume that obscured AGN generally have higher extinctions, AGN in older host galaxies will provide a majority of their flux in the same wavelength region as their galaxies. Balancing of the contribution of flux from these two components therefore leads to more complex degeneracies between the AGN luminosity, AGN extinction and host galaxy stellar mass and may therefore extend the distribution of AGN optical luminosity estimates. These effects can be seen in the 1-D posterior distributions of these properties shown in Fig. \ref{fig:corner}.  

\begin{figure}
    \centering
    \includegraphics[width=\linewidth]{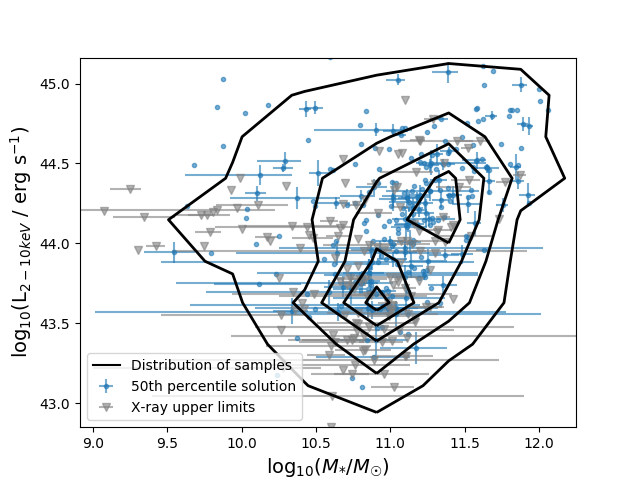}
    \caption{Host galaxy stellar mass MCMC inferences vs the measured X-ray luminosity from XMM-Newton. The blue points correspond to the median solutionsfor each object. For clarity, uncertainties limited to a third of the objects shown. For objects where log$_{10}(L_{2-10keV})$ is an upper limit, median solutions are shown in grey. Contours show all of the MCMC inferences for these objects.}
    \label{fig:lx_mass}
\end{figure}

\subsection{The stellar mass-AGN luminosity relation} \label{mstar}
To understand how host galaxy stellar mass inferences relate to AGN properties, we first look at the total distribution of derived host galaxy stellar masses, separated by the hardness ratio proxy for AGN type. These are shown in the right of Fig. \ref{fig:ebvhist}. Our findings show consistent host galaxy mass for unobscured and obscured AGN, having median mass values\footnote{The calculation of these values is discussed further in appendix \ref{appendix}} of log$_{10}(M_*/M_{\odot}$) = 10.88 $\pm0.09M_\odot$ and log$_{10}(M_*/M_{\odot}$) = 10.8 $\pm0.1M_\odot$ for unobscured and obscured objects respectively. Previous work by \cite{zou} and \cite{suh} have suggested a link between host galaxy stellar mass and AGN type. Their results varied, with the former finding unobscured AGN are typically found inhabiting less massive host galaxies than obscured AGN, and the latter finding the opposite. Both \cite{zou} and \cite{suh} do however differ in their methods of AGN type selection when compared to this work, instead using the presence of broad emission lines in the observed spectra to define type 1 AGN.

\begin{figure*}
    \centering
    \includegraphics[width=\textwidth]{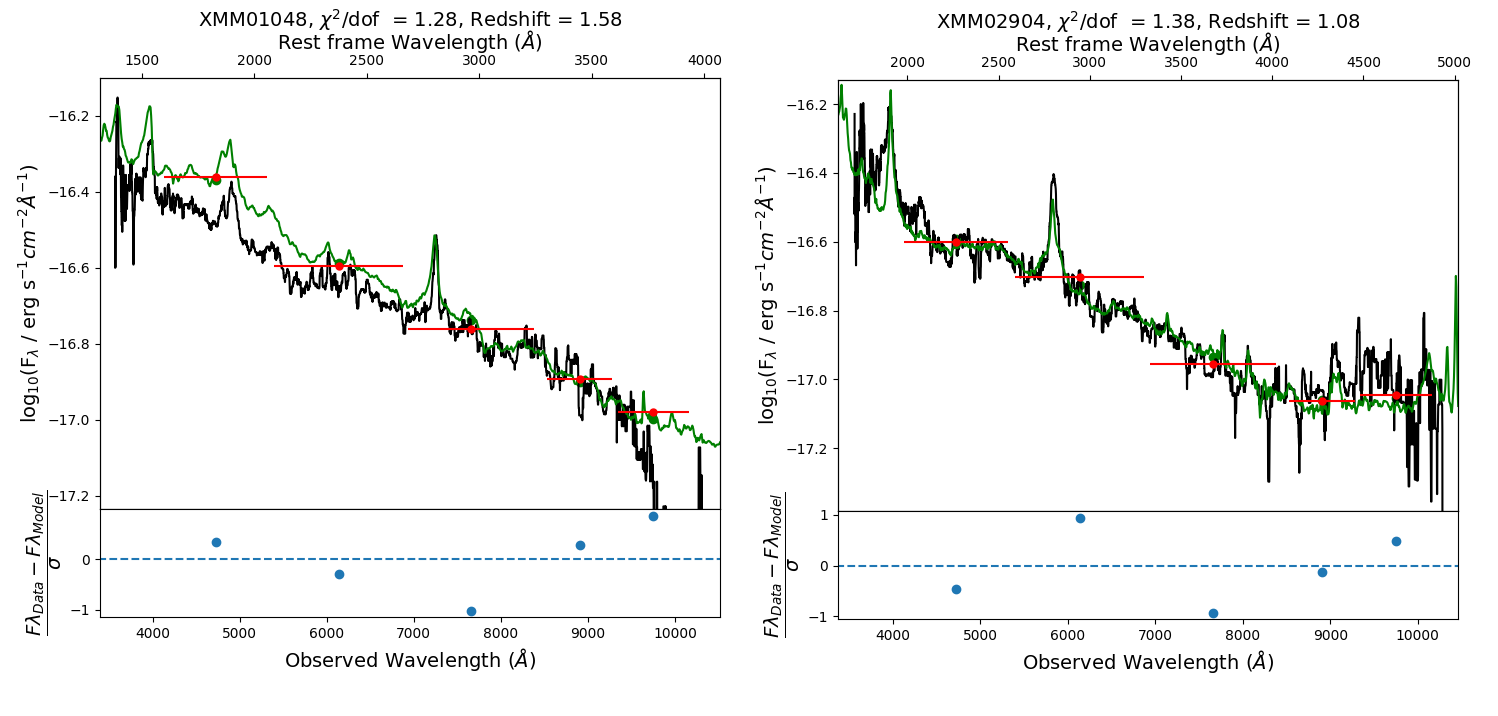}
    \caption{SDSS spectra (in black) comparison to the highest likelihood solution. The SDSS spectra is normalised to the best fit SED to have the same i-band flux. The figures show examples of an object that preferred weaker, more blue shifted lines on the left, and stronger, more symmetrical lines on the right, compared to the emission lines seen in AGN at z=2 and an average absolute magnitude of $M_i$ = -27. In the left figure, whilst the emission line strength appears to show good agreement between the width and strength of the lines observed, either AGN variability, or an offset in spectra calibration appears to change the continuum emission between the SDSS and best fit SED spectra.}
    \label{fig:sdss}
\end{figure*}

In order to further understand the underlying link between AGN and their host galaxies, we also looked at the inferred correlations between our AGN dust, optical, and X-ray luminosities with host galaxy stellar mass. We found that the strongest correlation of the AGN luminosities with stellar mass is with the X-ray luminosity, with a correlation coefficient of 0.15$\pm 0.03$. The uncertainty on this value was calculated using a Monte Carlo method using MCMC samples similarly to the calculation of property uncertainties as described in appendix \ref{appendix}, and shows a meaningful positive correlation between these AGN and host galaxy properties. This is slightly larger than the stellar mass-AGN optical luminosity, and stellar mass-hot dust luminosity correlations, which are 0.13 and 0.14 respectively. Focusing on the X-ray luminosity-stellar mass correlation, Fig. \ref{fig:lx_mass} shows the contours produced from our MCMC inferences comparing the measured X-ray luminosity with the stellar mass estimates, along with points corresponding to our median solutions. As in our previous analysis, both unobscured and obscured objects were found to show similar correlations, and as such are shown as a single distribution within Fig. \ref{fig:lx_mass}. A positive correlation between X-ray luminosity and host galaxy stellar mass has been also been found in previous work, such as \cite{Magliocchetti_2020}, and may provide insight into the nature of the AGN and host galaxy relations. 

 In terms of evolution with redshift, the median values of stellar mass MCMC inferences show no clear difference between galaxies at redshift < 1 compared to those at $z > 2$, with these values from our samples changing from log$_{10}(M_*/M_{\odot}$) = 10.83 $\pm ^{0.09}_{0.09}M_\odot$ to log$_{10}(M_*/M_{\odot}$) = 10.9 $\pm ^{0.1}_{0.1}M_\odot$. We see the uncertainty on stellar mass estimates increases with redshift. This could be due to the fact that whilst the stellar mass values don't decrease, high redshift objects are likely observed due to the presence of a high luminosity AGN, which can lead to more complex degeneracies that inflate the stellar mass uncertainties.
 
\subsection{AGN emission line properties} \label{emtype} 
A novel feature of the \citet{temple} quasar SED model is the incorporation of quasar emission line templates that reflect the full diversity of emission line strengths and morphologies seen in quasar spectra. The details of these emission lines are further described in Section \ref{AGN_properties}. A key question is whether these different emission line properties materially impact the broadband colours of quasars in such a way that the quasar emission line properties  can be inferred from SED fitting. In order to conduct this test, the SDSS Data Release 16 quasar catalogue \citep{Lyke:20} was matched to the XMM-SERVS sample of X-ray AGN to provide a sub-set of 408 AGN where emission line properties can directly be inferred from the spectra. As detailed in Section \ref{bestfd}, we fit SED models with the quasar emission line properties fixed to the default value, which represents the average emission line properties for luminous, high-redshift SDSS quasars \citep{temple}, as well as SED models where the emission line properties are a free parameter. The model with the lowest AIC is chosen as the best-fit model from all SED fits. 

We find that 61 of the 289 XMM-SERVS-SDSS matched catalogue AGN, with a best fit reduced $\chi^2 < 3$, prefer non-standard emission line properties. The average redshift of this sample has a higher mean redshift of $z = 1.86$, compared to $z = 1.46$ for the standard emission line sample. For the fixed emission model family fits, we find that 64\% have $z < 1.6$, which is only the case for 39\% of the free emission line fits.
This difference is expected, as for higher redshift objects, the strong emission line C{\textsc iv} present within the AGN SED is redshifted into the HSC g-band. For the lower redshift objects, this emission line is outside of the range fit by the SED, and thus information on its nature cannot be inferred for this sample. 

The right-hand plot in Fig. \ref{fig:sdss} shows an example of one non-standard emission line object, fit using stronger, more symmetric lines (emline\char`_type = 2.92) when compared to the emline\char`_type = 0 emission lines that correspond to the average SDSS AGN at $z = 2$ and average absolute magnitude $M_i$ = -27. The SDSS spectrum for the quasar is over-plotted and demonstrates excellent agreement with the emission line strength independently inferred from the photometry. Visual comparison of all of the best-fit SEDs for the 61 AGN with SDSS spectra that prefer non-standard emission line properties found that 66\% of the AGN show excellent agreement between the spectra and the spectral line strengths inferred from the photometry. 
For a further 25\% of AGN, the emission line strengths inferred from our SED fits are in reasonable agreement with the SDSS spectra but small differences in the continuum emission between the spectra and the best-fit SED are observed. These could arise for example due to quasar variability, or due to the typical uncertainties in the absolute flux calibration of the SDSS spectra. An example of such an object can be seen in the left-hand panel of Fig. \ref{fig:sdss}, which also features an AGN fit with weaker, more highly blueshifted emission lines when compared to the average emission lines seen in AGN at $z = 2$ and an average absolute magnitude $M_i$ = -27. Only 9\% of the AGN with non-standard emission line properties were inconsistent with the SDSS spectra. In these cases the best-fit SEDs were often composites of young star-forming galaxies with an obscured quasar, whereas the SDSS spectrum confirms the presence of a relatively unobscured AGN with broad emission lines. The final 6\% of objects had very low signal-to-noise SDSS spectra, precluding any firm conclusions regarding their nature. 

Having confirmed that emission line properties can be effectively inferred from broadband photometry, we now consider the multi-wavelength properties of these sources in more detail. From the total sample of 61 AGN with non-standard emission line properties, 61\% preferred stronger, more symmetric lines, with the remaining 36\% preferring weaker, more highly blueshifted lines relative to the T21 average quasar SED. The AGN with stronger, symmetric lines have an average 3000\AA\, luminosity of log$_{10}$(L$_{3000}$/erg s$^{-1}$) = 44.9$\pm^{0.2}_{0.2}$, similar to the average AGN luminosity for the sample best fit with the default emission line template in the \citet{temple} model of log$_{10}$(L$_{3000}$/erg s$^{-1}$) = 44.7$\pm^{0.1}_{0.1}$. On the other hand, the AGN that prefer weaker, more highly blue-shifted emission lines, indicative of line driven winds, have a higher average luminosity of log$_{10}$(L$_{3000}$/ergs$^{-1}$) = 45.2$\pm^{0.2}_{0.2}$. These results confirm the well-known Baldwin effect -- namely that the equivalent widths of strong emission lines in quasar spectra are anti-correlated with the AGN luminosity \citep{baldwin}. Confirmation of this result from broadband SED-fitting however, gives us an additional confidence that our SED fits are indeed able to correctly infer emission line properties.

We also consider the stellar mass of the AGN host galaxies as a function of their emission line properties. AGN with stronger, more symmetric lines have an average log$_{10}(M_*/M_{\odot}$) = 10.9$\pm0.1M_\odot$, which is consistent with the average for AGN best-fit by the default emission line template log$_{10}(M_*/M_{\odot}$) = 10.86$\pm0.08M_\odot$. AGN with weaker, more highly blueshifted lines tend to have lower host galaxy stellar masses of log$_{10}(M_*/M_{\odot}$) = 10.7$\pm0.3M_\odot$.

The ratio of the X-ray luminosity at 2keV and the UV luminosity at 2500\AA is often used to represent the hardness of the AGN ionising SED and can provide insight into the connection between the X-ray corona and the accretion disk of the AGN. We calculate log$_{10}(L_{2500\AA})$ using the the same approach as described in Section \ref{agnlum}. To find log$_{10}(L_{2keV})$, we use the hard (2-10keV), and soft (0.5-2keV) X-ray bands from \textit{XMM-Newton}. For each pair of robust X-ray measurements, we calculate a power law relationship for the flux density, and use this to k-correct these data to 2keV. This provides a distribution of photon indices, with an average value of $\gamma = 1.5$. This was used as the photon index for the objects in the sample where X-ray measurements only provided an upper limit in either the hard or soft X-ray bands, and as such individual photon indices could not be calculated. These values were used to calculate $\alpha_{ox}$. $\alpha_{ox}$ is defined as the slope of the power law between the X-ray and UV luminosities using the relation:

\begin{equation}
\alpha_{ox} = 0.384Log_{10}(L_{2keV} / L_{2500\AA})
\end{equation}

\begin{figure}
    \centering
    \includegraphics[width=\linewidth]{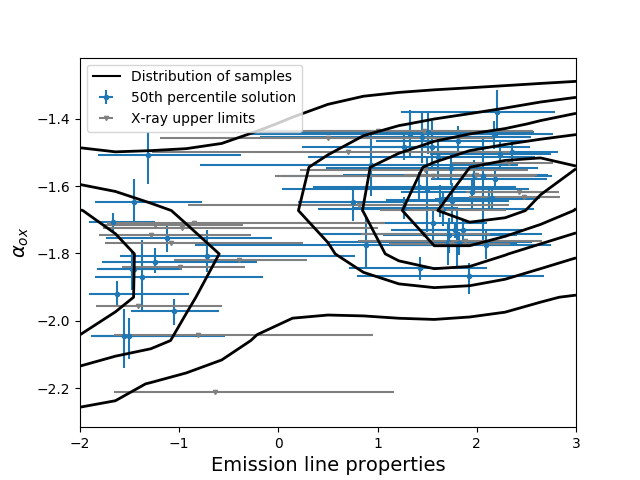}
    \caption{The X-ray to UV slope of the power law, $\alpha_{ox}$, vs the inferred AGN emission line properties. More negative values of emission line properties are indicative of line driven disc winds resulting in weaker, more blueshifted lines, whereas more positive values are stronger and more symmetric. The plotted sample does not include objects that preferred a fixed emission line within their fit, based on our AIC designation (such that the emission line properties value is fixed to 0).}
    \label{fig:aox}
\end{figure}

Our inclusion of emission line properties within the SED fits, and confirmation of their validity in Section \ref{emtype}, allows us to compare our inferred power law slope to the AGN emission line properties. Previous work, such as \cite{timlin_aox}, has found a significant correlation between $\alpha_{ox}$ and both the equivalent width and blueshift of the CIV emission line. Fig. \ref{fig:aox} shows the distribution of MCMC inferences for the sub-selection of 82 AGN+GAL model family objects that preferred non-standard emission line properties. It is important to note the bias in this sample, in that it only contains objects with a measurable difference in broadband flux as a result of the varying emission line properties when compared to the average $z = 2$, absolute magnitude $M_i$ = -27 AGN. In Fig. \ref{fig:aox} we see a positive correlation between $\alpha_{ox}$ and emission line properties for our inferences, with a correlation coefficient of 0.58. This suggests that the weaker blueshifted emission line objects, with smaller emission line equivalent widths, typically have softer AGN SEDs when compared to the stronger, more symmetric emission line objects. Previous studies \citep{timlin_aox, richards_aox} have found similar results using SDSS spectra. However, our work shows it is possible to see such a trend using exclusively photometry.

\section{Conclusions}
We performed SED-fitting to a spectroscopically confirmed sample of 711 high-redshift ($z > 0.7$) X-ray selected AGN in the XMM-SERVS survey using optical and near infrared photometric data from the HSC Deep/UltraDeep, VISTA VIDEO and \textit{Spitzer} SERVS surveys. 510 of these AGN were found to have reliable photometry across all 10 filters and a reduced $\chi^2 < 3$. We used the Aikake Information Criterion to classify these 510 AGN, and found 71 objects are galaxy-dominated, and 2 objects can be fit by a pure AGN template across the chosen wavelength range. 437 objects require both AGN and host galaxy components in the SED, thereby allowing the link between AGN and host galaxy properties to be investigated. We used a newly developed AGN SED model from \citet{temple} to characterise the AGN properties. The model incorporates a concise parametrisation of the variation in emission line properties across unobscured AGN, thereby allowing us to potentially infer AGN emission line properties from broadband photometry. In order to study the effects of bimodal or poorly converged solutions on our inferred properties, we investigate both the median solutions and full posterior distribution of MCMC inferences. These two methods are shown to provide consistent results for the properties investigated in this work.

\begin{itemize}
    \item 
    We find that the AGN X-ray luminosity is correlated with the 3000\AA \ luminosity inferred from SED-fitting in good agreement with previously known relations (e.g. \citealt{marconi}). These results demonstrate that robust AGN luminosities can be inferred from the SED fits.
\end{itemize}

\begin{itemize}
    \item Comparison with previous work found an offset of -0.23 dex between the AGN optical luminosity and hot dust luminosity at 3$\mu m$ from the relation found by \cite{jun_im}, possibly caused by the differences in the AGN templates used in these works. 
\end{itemize}

\begin{itemize}
    \item 
   We used the X-ray hardness ratio (HR) to split the AGN into X-ray obscured (HR > -0.2) and X-ray unobscured (HR < -0.2) AGN. This showed a consistent distribution for both obscured and unobscured in AGN luminosity. As might be expected for obscured AGN, an extended tail to larger E(B-V) was seen in the MCMC inferences, suggesting a generally higher extinction for obscured AGN.  We also find similar stellar masses for the sample of unobscured AGN (log$_{10}(M_*/M_{\odot}$) = 10.88 $\pm 0.09M_\odot$ and log$_{10}(M_*/M_{\odot}$) = 10.8 $\pm 0.1M_\odot$ for HR < -0.2 and HR > -0.2 AGN respectively). 
\end{itemize}

\begin{itemize}
    \item The ability to vary emission line properties in the T21 AGN template used in SED fitting allowed us to determine that for 18.8\% of the reduced $\chi^2 < 3$ sample, non-standard emission line properties were preferred over the average $z=2$, absolute magnitude $M_i$ = -27, AGN emission lines. By comparing to the SDSS spectra available for a subset of 82 AGN, we found that the emission line strengths inferred via SED-fitting to broadband photometry were broadly consistent with the results from spectroscopy for $\sim$91\% of the sample. These results highlight that the current generation of precision photometric datasets are able to infer emission line properties of broad-line AGN from photometry alone for a subset of AGN.
\end{itemize}

\begin{itemize}
    \item 
    We calculated $\alpha_{ox}$ based on the measured X-ray luminosity and the rest-frame UV luminosity inferred from our SED fits. We found a correlation between $\alpha_{ox}$ and the emission line properties inferred from photometry. This correlation showed that weaker, more blueshifted emission lines, indicative of line driven winds, were found to occur with softer $\alpha_{ox}$ slopes. In contrast, stronger, more symmetric emission lines preferred harder $\alpha_{ox}$ slopes. This is consistent with previous works \citep{richards_aox}, but hasn't previously been found using photometry alone.
\end{itemize}

The insights gained from our SED modelling could be used to aid in target selection for upcoming spectroscopic surveys such as 4MOST \citep{4most} and VLT-MOONS \citep{maiolino_2020}. SED fitting using the methods described in this paper will be applied to larger, non X-ray selected photometric samples to identify high luminosity AGN. Our ability to infer information on emission line morphology also allows for the selection of atypical AGN, such that in the future, further properties can be studied in greater detail. 

\section*{Acknowledgements}
AM acknowledges funding from a Royal Society studentship. MWA-W acknowledges support from the Kavli Foundation. MB acknowledges funding from The Royal Society via a University Research Fellowship. The authors would like to thank Paul Hewett and Matthew Temple for useful discussions. 
RM acknowledges ERC Advanced Grant 695671 QUENCH, and support from the UK Science and Technology Facilities Council (STFC). RM also acknowledges funding from a research professorship from the Royal Society. RB acknowledges support from an STFC Ernest Rutherford Fellowship [grant number ST/T003596/1].

\section*{Data availability}
The data used in this work may be shared on reasonable request to
the authors.

\bibliographystyle{mnras}
\bibliography{references} 

\appendix 
\section{Determining mean stellar masses for different populations  from MCMC samples} \label{appendix}
Our SED modelling has been undertaken independently for each object in this study, but it is also useful to determine properties of \textit{populations} of objects.  For concreteness, we will consider the case of inferring the mean stellar masses of the populations of AGNs with harder and softer X-ray emission, as discussed in Section 4.2. In principle, we could simultaneously investigate all objects with a \textit{hierarchical} model, assuming that each object's stellar mass is drawn from a Gaussian distribution for example, and sample for the population \textit{hyperparameters} (e.g., the mean and variance of the Gaussian distribution from which each object in a population might be drawn).

However, we have already obtained sampled realisations of population stellar masses from our independent MCMC analyses, by considering the set of samples from a given step of each object's Markov chain. We can therefore calculate the \textit{sample mean} $M_\textrm{pop,i}$ (and \textit{standard error on the sample mean}, $\sigma_\textrm{pop,i}$) for each $i$th MCMC step of all obscured or unobscured AGN, for example. This gives us a distribution of samples for the population mean stellar masses (i.e., the set of all $M_\textrm{pop,i}$), with uncertainties, that also implicitly propagates the uncertainties on the individual stellar masses from the SED fitting. In practice, we find that the standard error on the sample mean, $\sigma_\textrm{pop}$, is nearly the same for all MCMC steps, and we therefore quote the inference on the population stellar mass as the mean of the samples $M_\textrm{pop,i}$ with an uncertainty given by the square root of the sum of the variance of these samples and $\sigma^2_\textrm{pop}$.





\bsp	
\label{lastpage}
\end{document}